\documentstyle[preprint,prb,aps]{revtex}
\title{Energy cost associated with vortex crossing in superconductors}
\author{M.A. Moore}
\address{Department of Theoretical Physics, University of Manchester,\\
Manchester, M13 9PL, United Kingdom}
\author{N.K. Wilkin}
\address{Department of Physics,
University of Sheffield,\\ Sheffield, S3 7RH, United Kingdom.}
\date{29$^{\rm th}$ April 1994}
\begin{document}
\maketitle
\tighten
\draft
\begin{abstract}

Starting from the Ginzburg-Landau free energy of a type II
superconductor in a magnetic field we estimate the energy associated
with two vortices crossing. The calculations are performed by assuming
that we are in a part of the phase diagram where the lowest Landau
level approximation is valid. We consider only two vortices but with
two markedly different sets of boundary conditions: on a sphere and on
a plane with quasi-periodic boundary conditions.  We find that the
answers are very similar suggesting that the energy is localised to
the crossing point.  The crossing energy is found to be field and
temperature dependent -- with a value at the experimentally measured
melting line of $U_\times \simeq 7.5 k T_m
\simeq 1.16/c_L^2$, where $c_L$ is the Lindemann melting criterion parameter.
The crossing energy is then used with an extension of the Marchetti,
Nelson and Cates hydrodynamic theory to suggest an explanation of the
recent transport experiments of Safar {{\em et al.}\ }.

\end{abstract}
\label{vo}
\section{Introduction}
\label{vo:s:in}

Thermal fluctuations are widely believed to play an important role in
the physics of high temperature superconductors. In this paper we
consider a possible manifestation of these fluctuations --- that the
vortex lines will be able to cross through each other. We begin by
estimating the energy associated with such a process starting from the
Ginzburg-Landau free energy functional. We then analyse some of the
transport measurement data of Safar {{\em et al.}\ } \cite{Safar94},
which has connections with the flux line crossing energy, using a
modification of the Marchetti, Nelson and Cates hydrodynamic theory
for vortex motion \cite{Marchetti,Nelson93,Cates}.

The crossing energy calculation has previously been attempted within
London theory, an approximation which is valid when a typical vortex
separation is much greater than the vortex core size, and by
considering just two vortices in a fixed background
\cite{Sudbo,Wagenleithner,Carraro94}. However, London theory has an
unsatisfactory feature in that it requires a cut-off to be introduced
at the length scale of the core radius.  This is a difficulty as for
crossing the cores of the vortices need to overlap. We shall instead
provide an estimate of the crossing energy of two vortices using
Ginzburg-Landau (GL) theory, and the LLL (lowest Landau level)
approximation. Due to numerical difficulties our system is also
restricted to two vortices, but with two very different choices of
backgrounds: we have them moving on a sphere, and also on a plane with
quasi-periodic boundary conditions. In the planar case we consider a
unit cell commensurate with a triangular lattice ground state. We then
calculate the energy difference with respect to the perfect vortex
lattice. The results of these calculations are similar, suggesting
that the extra energy of two crossing vortices is localized to the
crossing region.

\section{Review of Landau-Ginzburg Theory Results}
\label{vo:s:rev}
To fix notations, we briefly describe anisotropic Landau-Ginzburg
theory for a superconductor, where $\psi$, the wavefunction is our
spatially dependent order parameter. For a fuller explanation and
justification see \cite{Wilkin,RT}.

We start with the free energy functional,
\begin{equation}
\label{vo:e:lg1}
\frac{F[\psi(r)]}{k_{B}T_{c}}=\int \mbox{d}^{3}r \left(\alpha(T)\,
|\psi|^2+\beta\,\frac{|\psi|^{4}}{2} + \sum_{\mu=1}^3
\frac{|(-i\hbar\partial_{\mu}-2e\mbox{\bf A}_{\mu})\psi|^{2}}{2
m_{\mu}}\right)+\frac{B^2}{2\mu_0}\: .
\end{equation}
Here $\alpha (T)$ is the temperature-dependent variable, $\beta$ is
the coupling constant, and $m_{\mu}$ is the effective mass. In the
cases we consider the masses in the $ab$-plane are taken as equal and
are denoted by $m_{ab}$, and the mass in the c-direction is written as
$m_c$. The temperature dependence of $\alpha(T)$ is taken to be
linear, $\alpha(T)= (T-T_{c}) \alpha'$. We also assume the lowest
Landau level (LLL) approximation, which is valid near the H$_{c_2}$
line. If we allow the variation in $B$ to be determined by the
equation $B=\mu_0 H_0 +(\mu_0 e \hbar)/(m_{ab}) \langle \left | \psi
\right |^2
\rangle$ (valid near to the H$_{c_2}$ boundary) then we can
write our temperature variable as $\alpha_{H}
=\alpha+e\mu_0 H \hbar/m_{ab}$. This is zero along the  $H_{c_{2}}$
line. A neater variable to work with is $\alpha_T$, which is dimensionless. It
is related to $\alpha_H$ by,
\begin{equation}
\label{vo:e:alph}
\alpha_H= \left ( \frac{\beta e \mu_0 H k_{B} T \sqrt{2 m_c}}{4 \pi
\hbar^2} \right ) ^{2/3}\alpha_{T}.
\end{equation}

The temperature dependence of the $\alpha_{T}$ variable is such that high
temperature is represented by $\alpha_{T} \rightarrow \infty$, low
temperature by $\alpha_{T} \rightarrow -\infty$ and $\alpha_{T}=0$
corresponds to being on the $H_{c_2}$ line. The above treatment is
valid where the LLL approximation can be trusted which probably
requires at the very least that $H > H_{c_2}/f$, where $f=3$ according to
Te\u{s}anovi\'{c} et al.\ \cite{Tesanovic}. The LLL approximation
allows us to write the order parameter as a set of orthonormal basis
functions, the form of these functions being determined by the
boundary conditions imposed to account for the finite size of the
system.

\section{Determination of the crossing energy}
\label{vo:s:det}
In this section we outline the method used for calculating the
crossing of two vortices within the GL framework, and using the LLL
approximation. Furthermore our entire system only contains the two
crossing vortices. For such a small system there is clearly a risk
that the crossing energy will be heavily dependent on the boundary
conditions used.  Hence we have carried out the calculation with two
different kinds of boundary conditions, firstly with the vortices
confined to the surface of the sphere and then on a plane filled with
unit cells with quasi-periodic boundary conditions. The results of our
two calculations yield remarkably similar results --- indicating that
the crossing energy is relatively insensitive to the presence of the
other vortices in the system.

\subsection{On a sphere}
\label{vo:s:det:s:sp}
We first consider the two vortices moving on a surface of a sphere,
a geometry that has been shown to reduce finite size effects in
numerical studies of superconductivity \cite{Jon0,Jon1} and the quantum Hall
effect \cite{Haldane}. The disadvantage of this approach is that the
sphere  is not commensurate with the triangular (Abrikosov) lattice
ground-state. However, as we will show later, the energy of our ground-state
is not very different from that of the triangular lattice.

On a  sphere we use the formalism of O'Neill and Moore
\cite{Jon1}. The starting line is the free energy functional given in
Eq.~(\ref{vo:e:lg1}). We then place a monopole at the center of the sphere,
which produces a radial magnetic field, satisfying $B 4 \pi R^2 =N
\Phi_0$, where $N$ is the number of vortices, which is two in our case. A
choice of vector  potential compatible with this condition is: $A_r=A
_{\theta}=0$, $A_{\phi}=B R \tan (\theta/2)$, in the usual spherical
polar coordinates.  The order parameter $\psi(\theta,\phi)$ can then
be expanded as eigenstates of the operator $(-i \hbar \nabla -2 e {\bf
A})^2/2m_{ab}$. Within the LLL the orthonormal set can be written as:
\begin{equation}
\label{vo:e:psi1}
\psi(\theta,\phi,h)=\sum_{m=0}^{N}v_{m}(h)\psi_{m}(\theta,\phi)
\end{equation} where, \begin{equation} \psi_{m}(\theta,\phi)=
k_{m}\sin^{m} (\theta/2)\cos^{N-m}(\theta/2)e^{im\phi}\;,\label{2}
\end{equation}
with $m=0, 1...N$; $k_{m}=[(N+1)!/4 \pi R^{2}
m!(N-m)!]^{1/2}$ and $h$ is the distance along the vortex line and
runs from $-\infty$ to $\infty$. The zeros of $\psi$ correspond to the
positions of the vortices at height $h$. The requirement for `crossing' is
that the two zeros  lie on top of each other at some value of $h$. We have
calculated the energy associated with the vortices being in their
equilibrium configuration at both ends and `crossing' at $h=0$. This
has been done by writing down the Euler-Lagrange equations associated with
two vortices and then solving with boundary conditions at $h=0$ and
$h=\infty$  compatible with the `crossing' requirement. (Technical
details of the calculation can be found in Appendix~\ref{vo:ap:s:sp}.)

To follow the above procedure we need to write the free energy in
terms of the coefficients $v_m$.
The general expression becomes \cite{Jon1},
\begin{eqnarray}
\frac{F[\{v_{m}\}]}{kT_c}&=&\int_{-\infty}^{\infty}\mbox{d}h\left [\,\sum_{m=0}
 (\frac{\hbar^2}{2m_c} \mid \frac{
\partial v_m}{\partial h}
\mid^2 +\alpha_{H}|v_{m}|^2)\right.\nonumber\\&&\mbox{}\left.+ \frac{\beta
B}{\Phi_0}\sum_{m,n,p,r=0}^{N}W(m+p,m,n)v_{m}v_{p}
v_{n}^{*}v_{r}^{*}\delta_{m+p,n+r}\:\right],
\label{vo:e:lg2}
\end{eqnarray}
where
\begin{eqnarray}
W(m+p,m,n)&=&\frac {2(N+1)^{2}}{N(2N+1)}
{}~~~\frac{f(m+p,m,n)~f(2N-m-p,N-m,N-n)}{f(2N,N,N)}\nonumber\\
{\rm and   }\ \ f(x,y,z)&=&x!/[(y!z!(x-y)!(x-z)!)^{1/2}2^{x+2}]\;.
\label{vo:e:w1}
\end{eqnarray}
{}From Eq.~(\ref{vo:e:lg2}) and Eq.~(\ref{vo:e:w1}) we find that the
free energy/per unit length in the z-direction for two vortices can be
written as:
\begin{eqnarray}
\label{vo:e:fe}
F & = &
  \int_{-\infty}^{\infty}\mbox{d}h
   \left [ \,\sum_{i=0}^2
     (\frac{\hbar^2}{2m_c} \mid \frac{ \partial v_i}{\partial h}\mid^2 +
      \alpha_H \mid v_i \mid ^2) +
     \beta B/\Phi_0 (0.45 |v_0|^4 +0.3 |v_1|^4 \right.
\nonumber\\&&\mbox{}+0.45 |v_2|^4 +0.9 |v_0|^2 |v_1|^2
                    +0.3|v_0|^2 |v_2|^2 +0.9|v_1|^2 |v_2|^2
\nonumber\\&&\mbox{}\left.+
                    0.3(v_0v_2v_1^{\ast} v_1^{\ast}  +
                    v_0^{\ast} v_2^{\ast}  v_1v_1) )
\vphantom{\sum_{i=0}^2}\right ].
\end{eqnarray}

The first task in evaluating the crossing energy $\Delta F$, is to
find the baseline, the free energy minimum of the ground state of the
system. It is easily found by minimisation of Eq.~(\ref{vo:e:fe}) to
be $-(\alpha_H^2 \Phi_0)/(1.2 \beta B)$ per unit length, as compared
with the Abrikosov ground-state of $-(\alpha_H^2 \Phi_0)/(\beta_a
\beta B)$.  Hence we have an effective $\beta_a=1.2$, which is a
reasonable approximation to the Abrikosov triangular lattice value of
$1.16$.  This solution corresponds to $v_0=\pm c$, $v_2= \mp c$, where
$c=\sqrt{|\alpha_H|\Phi_0/1.2 \beta B}$. To find the crossing energy
we next write down the Euler-Lagrange equations for the coefficients
$v_i(h)$. Solutions exist for which $v_1\equiv 0$ and in which
$v_0(-\infty)=-c$, $v_0(\infty)=c$, $v_2(-\infty)=c$, $v_2(\infty)=c$.
The crossing point then occurs at $h=0$ when $v_0=0$, which happens in
our notation to be when both vortices are at the south pole of the
sphere. (See Appendix \ref{vo:ap:s:sp} for details.)

The idea was to start from the crossing point and integrate forward
numerically along $h$ to the ground state configurations. Using energy
conservation and symmetry arguments at $h=0$ left us with one
remaining unknown parameter there. We then varied this parameter until
we found a value that allowed us to integrate forward and obtain our
equilibrium configuration. Classically this can be thought of as
finding the trajectory of a particle that is stationary at the maximum
in the potentials at (-c,c) to (c,c) , which moves as dictated by the
Euler-Lagrange equations. (See Fig.~(\ref{vo:f:pot}) for this
potential, and the actual trajectory taken.)

Having found a suitable trajectory, we could numerically integrate
along the path to find the difference in energy from the equilibrium
state. This left us with a prediction for the crossing energy of:
\begin{equation}
\label{vo:e:cr1}
\Delta F=\frac{1.3 \mid \alpha_H \mid^{3/2} \hbar
\Phi_0}{\beta B \sqrt{2 m_c}}= 0.33 \mid \alpha_T\mid^{3/2}\; k T.
\end{equation}

Our procedure could be generalized to handle several vortices. For two
vortices, use of symmetry and conservation arguments resulted in only
one adjustable parameter, but with three vortices (say) we would have
at least a three dimensional space to search. The problem rapidly
becomes completely intractable as the number of vortices increases and
hence we did not attempt to go beyond two.

\subsection{On a plane}
\label{vo:s:det:s:pl}

Given that we could not easily solve for more than two vortices on a
sphere, we instead looked at the same problem under very different
boundary conditions. Fortunately we obtain a very similar answer which
suggests that the crossing energy is determined mainly by the two
vortices which are being crossed. This second set-up was on the plane,
using the boundary conditions suggested by Kato and Nagaosa
\cite{Kato}. A rectangular unit cell is chosen with sides commensurate
with the formation of a triangular lattice of vortices, in effect
forcing the ground-state to be a triangular lattice. The cell also has
quasi-periodic boundary conditions such that the same motion of the
vortices is carried out in all of the unit cells simultaneously, which
at first sight looks rather unrealistic!

Within the LLL, and working in the gauge ($0,B x$)  the order parameter
can be written as,
\begin{equation}
  \psi (x,y,h)=\sum_{n=0}^{N-1} c_n(h) \phi_n(x,y)\:,
\end{equation}
where the $\phi_n$ are defined by (see Ref.~\onlinecite{Kato}):
\begin{equation}
\label{vo:e:phinn}
  \phi_n(x,y)\equiv
    \sqrt{\frac{1}{L_y \sqrt{\pi}l}}
      \sum_{m=-\infty}^{\infty}
        \exp \left[
          -i \left(\frac{2\pi l^2}{L_y}n+m L_x\right)\frac{y}{l^2} -
          \frac{1}{2 l^2}\left(x-\frac{2 \pi l^2}{L_y} n -m L_x \right)^2
        \right ]
\end{equation}
and $l^2=L_x L_y/(4 \pi)$. $L_x$ and $L_y$ are the sides of the
cell. The ground state is a triangular lattice if
$\sqrt{3} L_y=L_x$.
The periodicity of $\phi_n$ is such that
\begin{equation}
\phi(x,y+L_y) = \phi(x,y) \qquad {\rm and} \qquad \phi(x+L_x,y) =
\phi(x,y)\exp(-i\frac{L_x y}{l^2}).
\end{equation}

If we substitute the order parameter into Eq.~(\ref{vo:e:lg1}), for general
$N$ (number of vortices in the system) we obtain an expression for
$F[\{c_n\}]$. We write this using some auxiliary notation, explained
below;
\begin{eqnarray}
  F[\{c_n\}]	& = &
   \int_{-\infty}^{\infty} \mbox{d}h  \sum_{n=0}^{N-1}
     \left[ \left(
        \frac{\hbar^2}{2m_c}
          \left|\frac{ \partial c_n}{\partial h}\right|^2+
        \alpha_{H}|c_n|^2
       \right) +\mbox{   } \  			\right.		\nonumber\\
&& \left. \frac{\beta B}{4\Phi_0}
       \sqrt{\frac{L_x N}{L_y}}
          \frac{1}{(2 M+1)}
            \sum_I c_{n_1}c_{n_2}c_{n_3}^* c_{n_4}^*
              \exp\left( -\frac{\pi L_x}{2 N}(P^2+Q^2)\right)  \right ].
\label{vo:e:lg3}
\end{eqnarray}
The sum indicated by the symbol $\sum_I$ runs over the eight variables
$n_1,\ldots n_4$ and $m_1,\ldots m_4$, subject to the requirements that
\begin{equation}
  n_1 + n_2 + N(m_1 + m_2) = n_3 + n_4 + N(m_3 + m_4)
\end{equation}
\begin{equation}
  0 \le n \le N-1	\qquad \mbox{ and } -M \le m \le M
\end{equation}
The symbols $P$ and $Q$ are defined as
\begin{eqnarray}
  P & = & n_1 - n_2 + N(m_1 - m_2)	\\
  Q & = & n_3 - n_4 + N(m_3 - m_4)\:.
\end{eqnarray}

If we now consider only two vortices and evaluate the sums, we are left
with the equation for the free energy in the form:
\begin{eqnarray}
\label{vo:e:lg4}
F&=& \int_{-\infty}^{\infty}
\left [\sum_{i=0}^1 (\frac{\hbar^2}{2m}
\mid \frac{ \partial c_i}{\partial h} \mid^2 +
\alpha_H \mid c_i \mid ^2) +\beta B/\Phi_0 (
p(|c_0|^4 +|c_1|^4) \right.\nonumber\\&&\mbox{} \left. +
q|c_0|^2|c_1|^2 + s(c_0^2c_1^{{\ast}^2} +c_0^{{\ast}^2}
c_1^2 ))\vphantom{\sum_{i=0}^1}\right],
\end{eqnarray}
where $p=0.465337,q=0.245010$ and $s=0.008063$.
The ground state energy per unit length is $F=-\alpha_H^2 \Phi_0/\beta_a
\beta B$
with $\beta_a=1.15956$. This is the triangular lattice result, accurate to five
decimal places, the error occurring in the evaluation of the sums in
$\phi_n$.

We now outline the calculation required to evaluate the
crossing energy on the plane, leaving the details for
Appendix~\ref{vo:ap:s:pl}. Symmetry suggests that a
suitable place for the zeros of the vortices to be superposed is at
the center of the unit cell.  To achieve this we imposed the
condition,
\begin{equation}
c_0(0) \phi_0\left(\frac{L_x}{2},\frac{L_y}{2}\right)
+c_1(0)\phi_1\left(\frac{L_x}{2},\frac{L_y}{2}\right)=0.
\end{equation}
The equilibrium configuration can be satisfied with $c_0(-\infty)=c$,
$c_1(-\infty)=-c$, $c_0(-\infty)=c$, $c_1(-\infty)=c$ where we know
$c=\sqrt{|\alpha_H| \Phi_0/\beta_a \beta B}$ by energy arguments. By
imposing `energy conservation', and recognising that we also have
`angular momentum conservation', in the dynamical analogue of a
particle moving in a potential it is possible to reduce our set of
Euler-Lagrange equations again to solving for just one unknown
parameter. We can then proceed in a similar manner to the case of the sphere,
although in this case we have numerical problems in reaching the
equilibrium condition and the energy we evaluate has a residual
kinetic energy which is 5\% of the maximal kinetic and 2\% of the
total energy, and hence an error of $\sim$5\% on the crossing
energy. The final value we calculate is very similar to that for the
sphere,
Eq.~(\ref{vo:e:cr1}),
\begin{equation}
\Delta F=\frac{1.46 \mid \alpha_H \mid^{3/2} \hbar \Phi_0}{\beta B
\sqrt{2 m_c}}= 0.37 \mid \alpha_T\mid^{3/2} \; k T.
\end{equation}

It should be remembered when comparing the two calculations of the
crossing energy that we already had a 4\% difference in the
ground-state energies for the different boundary conditions.

\section{Crossing Angle}
Previous calculations of the crossing energy within the London regime
by Obukhov and Rubinstein \cite{OR}  and Nelson \cite{Nelson93} have
considered a mechanism for crossing in which at the crossing point the
the configuration of the vortices is chosen for simplicity to be that shown in
Fig.~\ref{vo:f:nelson}. The crossing angle $\theta_c$ depends on the
anisotropy of the sample, $\theta_c\simeq \arctan
(\sqrt{m_c/m_{ab}})$, and is $45^\circ$ for an isotropic
superconductor.

In our calculation it was not necessary to impose the configuration of
the vortices at the crossing point, but in the planar case it can be
extracted from the solution to the Euler-Lagrange equations. (For the
case of the sphere it is difficult to decide how the angles map to
those of the flux lines physically crossing.) It should
be noted that the lowest energy configuration for the vortices to
cross may differ between the London and LLL regimes as in the latter
the variation of the $B$-field is not included.

On the plane, we know the positions of the zeros at all heights is
given by,
\begin{equation}
\label{vo:e:cra1}
c_0(h) \phi_0 (X,Y) +c_1(h)\phi_1(X,Y) \equiv 0,
\end{equation}

where $c_0(h)$ and $c_1(h)$ are
known from solving the Euler-Lagrange equations. In order to
investigate the behaviour of the flux lines near the crossing point we
consider deviations $(x,y)$ from the position at $h=0$,
$X=L_x/2,Y=L_y/2$, by considering the differential of
Eq.~(\ref{vo:e:cra1}) with respect to $h$; \begin{equation}
\overbrace{(\partial_h c_0) \phi_0  +(\partial_h c_1) \phi_1}^{\alpha}
+(\partial_h x) \overbrace{(c_0 \partial_x \phi_0 +c_1 \partial_x
\phi_1)}^{\beta}   +(\partial_h y) \overbrace{(c_0 \partial_y \phi_0
+c_1 \partial_y \phi_1)}^{\gamma}=0. \end{equation} We find that at
the crossing point $\alpha=0.565 i$ and that $\beta=5.26(x- i y)$ and
$\gamma=-5.26(ix+y)$. Hence solving for the two equations for the real
and imaginary parts we find that in the vicinity of the crossing point
$x=y=\sqrt{0.11 h}$.

This implies that the vortices save energy on close approach by
allowing their core energies to cancel (see
Fig.~\ref{vo:f:vonikki}). The vortices also approach not along the
body diagonal of the unit cell but at $45^{\circ}$ such that there is
a twist along the length of the vortices as they approach.

\section{Comparison with London theory results}
In order to allow comparison between our results for the crossing
energy and the estimates for crossing within the London regime, it is
simplest to re-express our crossing energy in terms of the Lindemann
melting criteria. This criteria suggests that when the thermally
induced spatial fluctuations in the positions of the flux lines
$\langle {\bf u}_{th}^2\rangle$ become of the order of $c_L ^2 a_0^2$
the lattice will melt. The Lindemann number can be obtained indirectly
from neutron scattering and estimates range from $0.1$--$0.4$.

By writing the Lindemann criterion in the form suggested by Moore
\cite{Moore89},
\begin{equation}
\langle {\bf u_{th}}^2\rangle\simeq
\frac{kT}{4\pi\sqrt{\rho_s c_{66}}}=c_L^2 a_0^2\,,\qquad
a_0^2=\frac{\Phi_0}{B}
\end{equation}
and using the formulae for
\begin{equation}
\label{vo:e:c66}
\rho_s=\frac{|\alpha_H| \hbar^2}{m_c \beta \beta_a}, \qquad
c_{66}=\frac{0.24 |\alpha_H|^2}{\beta_a^2 \beta}\:,
\end{equation}
which are determined using elasticity theory, \cite{Moore92} we find that
$c_L^2=3.13/|\alpha_T|^{3/2}$ for $|\alpha_T|$ at the melting line.
Hence, using the result for
$U_\times\simeq 0.35 |\alpha_T|^{3/2} kT$, we find that $U_\times /k T
\simeq 1.1/c_L^2$  (leading us to estimate $c_L \sim 0.4$). Nelson
\cite{Nelson93} has estimated in the London regime
that
\begin{equation}
\frac{U_\times}{k_B T}\simeq 2(\sqrt{2} -1)
\sqrt{\frac{m_{ab}}{m_c}}\, a_0 \, \ln \left[\kappa_{ab}\right] \;
\left(\frac{\Phi_0}{4 \pi \lambda_{ab}} \right)^2\;\sim\;
\frac{0.75}{c_L^2},
\end{equation}

which is of the same order of magnitude as our LLL estimate. A more
recent estimate, also in the London regime, by Carraro and Fisher
\cite{Carraro94} leads to a result of $U_\times /k T \simeq 0.24/c_L^2$,
which is rather lower than our result. There is of course no reason to
expect that the results for the crossing energy in the LLL and the
London regime to be identical.

\section{Discussion}

Having estimated the crossing energy we now consider its relevance
within the flux liquid phase of the high temperature
superconductors. In order to understand the magnitude of the crossing
energy in terms of physical quantities we first estimate its value in
terms of $k T_M$ at the melting line found experimentally by
Worthington {\em et al.} \cite{Worthington} and Safar {\em et
al.}\cite{Safar93}. (Providing we accept that the melting line can be
equated with the irreversibility line in magnetisation measurements.) The
latest data from Safar {\em et al.} has been taken in fields up to 16
T. Their results show a first order transition for fields below $\sim
10$T and a continuous transition (which is no longer a melting
transition) for larger fields. In order to consider the magnitude of
the crossing energy in the vicinity of the transition we need an
estimate of $\alpha_T$. From previous work
\cite{Wilkin2} we have a crude estimate of $\alpha_T \sim -8$ based on the
melting line shown in the Worthington {\em et al.} \cite{Worthington}
data.  The latter data is taken in the region now associated with the
first order transition but the value of $\alpha_T$ should not change
appreciably with increasing field. An alternative estimate \cite{HFL1}
from the analysis of theoretical specific heat data suggests
$\alpha_T\sim -7$, although their argument is somewhat suspect
\cite{Wilkin2}. If we believe in this magnitude of $\alpha_T$ then we
have $U_\times/k T\simeq 7.5 $ (for $|\alpha_T|=8$) in the region of
the phase diagram being investigated.

Previously Marchetti, Nelson and Cates \cite{Marchetti,Nelson93,Cates}
have shown using a hydrodynamic theory that the effects of a twin
boundary or similar pinning surface whose normal is perpendicular to
the field act over a characteristic length $\delta_{ab}$, when the
magnetic field is in the $c$-direction. This characteristic length is
effectively the distance between line crossing events in the
$ab$-plane. Cates \cite{Cates} found that $\delta_{ab}\simeq a_0 \exp
[U_\times/2 kT]$, where $a_0$ is the spacing between vortex lines. By
extending Cates argument we now deduce the length scale $\delta_c$,
relevant for velocity gradients varying along the field direction,
which we believe to be the important length scale in the transport
measurements of Safar {{\em et al.}\ } \cite{Safar94}. The length
$\delta_{ab}$ is the typical transverse distance between crossing
events, and can be written as
\begin{equation}
\delta_{ab}^2=l_p^2 \frac{\delta_c}{l_p}
\end{equation}
where this is just a `random walk' of $\delta_c/l_p$ steps. $l_p$ is a
persistence length, as described by Cates, and is related to $l_e$,
the distance along a flux line which one has to travel before
encountering another flux line by $l_e l_p =a_0^2$. The resulting
distance between crossing events in the c-direction is now given by:
\begin{equation}
\delta_c \simeq l_e e^{U_\times/kT}.
\end{equation}
It is clear that when $\delta_c$ becomes equal to the thickness of the
sample (or larger) then the vortex motion at the top face of the
sample will be strongly correlated with that at the bottom face of the
sample.  Experimentally the degree of correlation between the motion
of vortices at the top and bottom surfaces of the sample has been
measured by Safar {{\em et al.}\ } \cite{Safar94} on YBCO crystals and
has been analysed by Huse and Majumdar \cite{Huse93}.  The
experimental set-up involves applying a magnetic field along the
c-axis and then injecting a transport current in the top a-b plane
along the a-axis. The voltages in the top and bottom faces are then
measured and the maximum field temperature combinations above which
$V_{top}=V_{bot}$ are recorded. If the vortices are not readily able
to cross through each other then the voltages on the two faces due to
the Lorentz force induced on the top surface should be the
same. However, once $\delta_c=L$, the size of the sample, line
crossing will allow the vortices to remain pinned on the bottom face
whilst they continue to move due to the Lorentz force on the top
face. The mechanism for this procedure can be seen in
Fig.~\ref{vo:f:volt}. Initially, the flux line labeled `A-B' is in
front of the line `C-D' and the top of line `A' is subject to the
Lorentz force exerted by the transport current. However, line `C-D' is
pinned along its entire length and is unmoved by the Lorentz force. If
crossing is energetically favourable then when end `A' encounters end
`C' the two lines will cut and re-combine as `A-D' and `C-B' resulting
in a net movement of vortices, and hence a voltage in the top
surface. There is no net flux line movement and hence no voltage in
the bottom surface.

Using the results from Safar {{\em et al.}\ } and our value of the crossing
energy we find that $\delta_c =2.5 l_e \exp[U_\times/kT]$ is a good
fit to the data for the $L=30\mu$m sample. The value for $l_e$ was
estimated by assuming that the persistence length $l_p$ would be of
the order of the spacing between the Cu-O planes in YBCO. Considering
that $l_e$ is only an order of magnitude estimate, this mechanism is a
possible explanation of the results. We also find that for a given
magnetic field the point at which $V_{top}=V_{bot}$ decreases, with
increasing thickness of sample. This is in qualitative agreement with
experimental data at 1 T. However, the quantitative comparison is
poor, which is not surprising as 1 T is outside the expected region of
validity of the LLL approximation.

Clearly, this is a first estimate of the crossing energy, and a more
complete calculation will consider a larger number of vortices, as
well as allowing for fluctuations of the vortices. This calculation
has also indicated that the length scales in the problem are of the
order of the Cu-O layer spacing indicating that an investigation of
crossing within a layered structure is called for, rather than as here
just within the continuum GL theory.

We would like to thank Hugo Safar for the use of his experimental data
and Michael Cates and William Barford for many useful discussions.

\newpage
\appendix
\section{Details of the crossing energy calculations}
\label{vo:ap}  In this appendix
we give a more detailed description of the procedure used to calculate
the crossing energy on both the sphere and the plane. The problem we
had to solve was essentially a classical mechanics problem of finding
a trajectory for a ball to roll from one metastable position on one
local maximum to another at the same height, with a set of conditions
to control the velocity and position at the central point of the
trajectory.

\subsection{On a sphere}
\label{vo:ap:s:sp}
The first simplification we used on working on the sphere was to
invoke a reduced set of the variables $\{v_m\}$. We claimed in
\S\ref{vo:s:det:s:sp}
 that this subset was sufficient to yield the correct
solutions --- the justification is quite simple although algebraically a
little heavy.

We begin with the free energy in terms of the full set of $\{v_m\}$, as in
Eq.~(\ref{vo:e:lg2}), and the order parameter $\psi$ as defined in
Eq.~(\ref{vo:e:psi1}).
We then express $\psi$ in terms of spinor variables. In the case of
two vortices, N=2 this is simply:
\begin{equation}
\label{vo:e:psi2}
\psi(\theta,\phi)=A (w u_1-u w_1)(w u_2-u w_2)\:,
\end{equation}
where $u=\cos(\theta/2)e^{-i \phi/2}$,$w=\sin(\theta/2)e^{i \phi/2}$
and $A=r e^{i f}$ is a complex variable. The positions of the vortices are
described by the zeros of $\psi$ which occur when $u=u_i$ and
$w=w_i$. Hence we now know the positions of the vortices in terms of the
polar angles $\theta_i$  and $\phi_i$. By comparison of
Eq.~(\ref{vo:e:psi1})
and Eq.~(\ref{vo:e:psi2}) we can find the relationship between the $\{v_i\}$
and $A,\{u_i,w_i\}$ such that we can write Eq.~(\ref{vo:e:lg2}) in terms of
the spinor variables. A more coordinate-independent variable than the
individual positions of the vortices is the separation between them, a
measure of which is the scalar product $n$ of their positions. In
terms of the spinor variables  $n$ can be written as
$n=\cos(\theta_1)\cos(\theta_2) + cos(\phi_1-\phi_2) \sin(\theta_1)
\sin(\theta_2)$.  Inspection of the potential energy written in
terms of the spinor variables shows it can easily be re-written in
terms of $n$ and $r$:
\begin{equation}
\label{vo:e:potential}
V=\frac{r^2}{4}(3+n) +\frac{r^4}{160}(39+30n+3n^2).
\end{equation}
The kinetic energy is somewhat more complicated. However the terms
that cannot be written in terms of $n$ and $r$ can be written as sums
of squares. Hence they are effectively conserved momenta, whose
contribution will be zero when we look for the minimum energy solution.
The remaining relevant terms in the kinetic energy are then:
\begin{equation}
T=\frac{(3+n)}{4}(\partial_h r)^2+
\frac{r}{4}\,(\partial_h r) \, (\partial_h n)
+\frac{r^2}{16(1+n)}(\partial_h n)^2.
\end{equation}
If we then derive the Euler-Lagrange equations for this new system we
find they map back to just the real parts of $v_0$ and $v_2$. Hence
the set of variables we have chosen to work with is sufficient. We
should perhaps point out that there is good  reason for working with
the less intuitive variables $v_0$ and $v_2$--- the Euler-Lagrange
equations are then far simpler.

We are now in a position to calculate the crossing energy. We
start by writing down the Euler-Lagrange equations,
\begin{eqnarray}
\frac{\hbar^2}{2m_c} \partial_h^2 v_0&=& v_0 \left(-\mid
\alpha_H \mid  + \frac{\beta B}{\Phi_0} \, (0.45 v_0^2 + 0.3 v_2^2 )\right)
\nonumber\\
\frac{\hbar^2}{2m_c} \partial_h^2 v_2&=& v_2 \left(-\mid
\alpha_H \mid  + \frac{\beta B}{\Phi_0} \, (0.45 v_2^2 +0.3 v_0^2 )\right).
\end{eqnarray}
Then the total energy, $T+V=\Phi_0|\alpha_H|^2/(1.2 \beta B)$, does
not vary with $h$.  This is an initial value problem so we need to
know the set of \{$v_0,v_2,\partial_h v_0,\partial_h v_2$\} at h=0,
where the vortices cross. Examination of the boundary conditions
showed that $v_0$ was antisymmetric and $v_2$ symmetric about h=0 (see
Fig.~\ref{vo:f:v0v2}) and as a consequence $v_0=0$ and $\partial_h
v_2=0$ at $h=0$. Then, by conservation of energy, we knew that
$\partial_h v_1$ is a function of $v_2$. Hence we just searched the
one parameter space $v_0$ until we found a solution of the equations
whose end-point was compatible with the ground-state configuration.
The value of $h$ at which the vortices became straight again is $h\sim
6 \hbar/\sqrt{2 m_c\alpha_H}$. That is the crossing was completed on a
length scale $l_c$ of the order of the conventional phase correlation
length $\xi_c$, and $l_c\simeq 12 \xi_c$, see
Fig.~\ref{vo:f:vonikki}. In practice this length scale will be of the
order of the spacing of the Cu-O planes, indicating the need to go
beyond the continuum GL approach of this paper.

\subsection{On a plane}
\label{vo:ap:s:pl}
The procedure for calculating the crossing energy on a plane is very
similar to that just described for a sphere. Hence we will consider it
only briefly. In this case no attempt was made to reduce the set of
variables, so we had as our starting point a set of four coupled
non-linear Euler-Lagrange equations,
\begin{eqnarray}
\partial_h^2 r_0&=&r_0\left(2 p r_0^2 +r_1^2\left(q
	+2s \cos(2 \phi_0-2\phi_1)\right)  +(\partial_h\phi_0)^2
-1\right)\\
                 \partial_h^2 r_1&=&r_1\left(2 p r_1^2 +r_0^2\left(q
	+2s \cos(2\phi_0-2\phi_1)\right) +(\partial_h\phi_1)^2
-1\right)\\
	         r_0 \partial_h^2 \phi_0&=&-2s r_0 r_1^2
\sin(2\phi_0-2\phi_1)
 -2 (\partial_h r_0)( \partial_h \phi_0)\\
r_1 \partial_h^2\phi_1&=&\;\;\; 2s r_1 r_0^2 \sin(2 \phi_0- 2\phi_1)
-2 (\partial_h r_1)(\partial_h\phi_1)\:,
\end{eqnarray}
where $c_0=r_0 e^{i \phi_0}$ {\it etc.}, and $p,q,s$ are as defined for
Eq.~(\ref{vo:e:lg4}). Written in this form, we can see by inspection,
that as well as conservation of energy we also have conservation of
`angular momentum' as $\partial_h(r_0^2 \partial_h \phi_0+r_1^2 \partial_h
\phi_1)\equiv 0$. Moreover, because the boundary conditions at $\pm
\infty$ require there to be no kinetic energy, the angular momentum
must be equal to zero. So far we have a general set of equations
describing the motion of the two vortices within the unit cell, but we
need a further constraint to make the vortices cross.  Having decided
that a suitable place for the vortices to meet would be the center of
the unit cell, we find that we need to satisfy the condition,
$c_0(0)\psi_0(L_x/2,L_y/2)+ c_1(0)\psi_1(L_x/2,L_y/2)=0$. Evaluation
of the $\psi$'s shows that we require $r_1=0.13165 r_0$ and
$\phi_0=\phi_1=0$ at $h=0$. In fact the other obvious places for the
vortices to cross, $(\pm L_x/2,0)$ and $(0,\pm L_y/2)$ yield
equivalent constraints.  Through symmetry arguments we find that
$\partial_h c_0=\partial_h c_1=0$ at $h=0$.

Using all of these facts leaves us again with only one free parameter
in our initial conditions. Finding a solution which reached the
necessary end-point proved more tricky this time, and we were unable
to produce a solution which bettered having 5\% of the maximal  kinetic
energy (2\% of the total energy) in `equilibrium', and correspondingly
the equilibrium separation of the vortices is only accurate to 2\%.
Hence an estimate of the error on the crossing  energy of 5\% would
seem reasonable.

\newpage
\begin{figure}
\caption{Contour plot of the `potential' as described in
Eqn.~(\protect\ref{vo:e:potential}) for the spherical boundary
conditions problem. The solution to the Euler-Lagrange equations which
corresponds to the vortices crossing is that linking the points `A'
and `B', and $c$ is as defined in the text.}
\label{vo:f:pot}
\end{figure}
\begin{figure}
\caption[Ansatz configuration for vortices crossing in the London
regime]{Ansatz configuration for vortices crossing in the London
regime used in Ref.~\protect\onlinecite{Nelson93}, where
$\gamma=(m_c/m_{ab})^{1/2}$.
}
\label{vo:f:nelson}
\end{figure}
\begin{figure}
\caption{Configuration for vortices crossing within the LLL regime, $l_c$ is
the length over which the crossing takes place.}
\label{vo:f:vonikki}
\end{figure}
\begin{figure}
\caption{Possible vortex motion in the Safar {{\em et al.}\ } experiment.
Initially (a) line `A-B' is in front of C-D and subject to a Lorentz
 force at end `A', and line `C-D' is pinned. Crossing enables end `A'
 to carry on moving, (b), resulting in a net voltage difference
 between top and bottom of the sample.}
\label{vo:f:volt}
\end{figure}
\begin{figure}
\caption[Solutions of the Euler-Lagrange equations on the sphere] {The
functions $v_0(h)$ (antisymmetric)and $v_2(h)$(symmetric)  as
calculated by solving the Euler-Lagrange equations. $v_0$ and $v_2$
are in units of `$c$' and $h$ is in units of $\hbar/\sqrt{2
m_c\alpha_H}$, the c-axis correlation length. }
\label{vo:f:v0v2}
\end{figure}

\end{document}